\documentstyle[bbm,amsfonts,epsfig,12pt]{article}

\newcommand{\smallfrac}[2] {\mbox{$\frac{#1}{#2}$}}
\newcommand {\eqref} [1] {(\ref {#1})}

\newcommand {\slsh} [1] {\not{\hbox{\kern-2pt${#1}$}}}

\newcommand{\drawsquare}[2]{\hbox{%
\rule{#2pt}{#1pt}\hskip-#2pt%  left vertical
\rule{#1pt}{#2pt}\hskip-#1pt%  lower horizontal
\rule[#1pt]{#1pt}{#2pt}}\rule[#1pt]{#2pt}{#2pt}\hskip-#2pt%upper horizontal
\rule{#2pt}{#1pt}}% right vertical

\newcommand{\Yfund}{\raisebox{-.5pt}{\drawsquare{6.5}{0.4}}}%  fund
\newcommand{\Ysymm}{\Yfund\hskip-0.4pt%
                     \Yfund}%  symmetric second rank
\def\symm{\Ysymm}

\def\drawbox#1#2{\hrule height#2pt
         \hbox{\vrule width#2pt height#1pt \kern#1pt
               \vrule width#2pt}
               \hrule height#2pt}

\def\Fund#1#2{\vcenter{\vbox{\drawbox{#1}{#2}}}}
\def\Asym#1#2{\vcenter{\vbox{\drawbox{#1}{#2}
               \kern-#2pt       % line up boxes
               \drawbox{#1}{#2}}}}

\def\fund{\Fund{6.4}{0.3}}
\def\asymm{\Asym{6.4}{0.3}}

\newcommand {\beq} {\begin{equation}}
\newcommand {\eeq} {\end{equation}}
  \newcommand {\ber}{\begin{eqnarray*}}
  \newcommand {\eer} {\end{eqnarray*}}
\newcommand {\bea}{\begin{eqnarray}}
  \newcommand {\eea} {\end{eqnarray}}

\newcommand{\Dslash}{\,{\raise.15ex\hbox{/}\mkern-12mu D}}

\begin{document}

%%%%%%%%%%%%%%%%%%%%%%%%%%%%%%%%%%%%%%%%%%%%%%%%%%%%%%%%%%%%%%%%%%%%%%%%%%%%%%%

\begin{titlepage}

\begin{center}
\vspace{1in}
\large{\bf The Conformal Window from the Worldline Formalism}\\
\vspace{0.4in}
\large{Adi Armoni}\\
\small{\texttt{a.armoni@swan.ac.uk}}\\
\vspace{0.2in}
\large{\emph{Department of Physics, Swansea University}\\ 
\emph{Singleton Park, Swansea, SA2 8PP, UK.}\\}
\vspace{0.5in}
\end{center}

\abstract{We use the worldline formalism to derive a universal relation for the lower boundary of the conformal window in non-supersymmetric QCD-like theories. The derivation relies on the convergence of the expansion of the fermionic determinant in terms of Wilson loops. The expansion shares a similarity with the lattice strong coupling expansion and the genus expansion in string theory. Our result relates the lower boundary of the conformal window in theories with different representations and different gauge groups. Finally, we use SQCD to estimate the boundary of the conformal window in QCD-like theories and compare it with other approaches.}

\end{titlepage}

%%%%%%%%%%%%%%%%%%%%%%%%%%%%%%%%%%%%%%%%%%%%%%%%%%%%%%%%%%%%%%%%%%%%%%%%%%%%%%%%
\section{Introduction}

 Consider an $SU(N_c)$ gauge theory with $N_f$ massless Dirac fermions in the fundamental representation. When $N_f ^\star < N_f < {11\over 2} N_c$ the theory flows in the IR to a fixed point. This range of $N_f$ is often called the conformal window. A conformal window is expected to occur in other theories based on different gauge groups and matter representations. 

The precise value of $N_f ^\star$ requires a knowledge of the vacuum structure of the theory. Due to holomorphicity it can be calculated in SQCD \cite{Seiberg:1994pq}. The same value $N_f ^\star = {3\over 2} N_c$ holds also in a non-supersymmetric large-$N$ ``orientifold'' version of SQCD \cite{Armoni:2008gg}. In QCD (or QCD-like theories) there were various attempts to estimate the conformal window \cite{Appelquist:1996dq,Miransky:1996pd,Appelquist:1988yc,Appelquist:1998rb,Gardi:1998ch,Appelquist:1999hr,Dietrich:2006cm,gies,Ryttov:2007cx,Poppitz:2009uq}. Most of the methods assume that confinement is linked with chiral symmetry breaking, due to an argument by Casher \cite{Casher:1979vw}. An interesting exception is ref.\cite{Poppitz:2009uq}, where the authors consider the theory on $R^3 \times S^1$ and by exploiting the index theorem \cite{Poppitz:2008hr} argue that the theory is confining as long as there is a mass gap in the gauge sector. We will use a similar criterion in the present paper. Finding the value of $N_f^\star$ in various QCD-like theories is under intensive investigation by several lattice groups \cite{lattice}. This topic is nicely reviewed in a recent talk by Peskin \cite{peskin}.

In this short note we propose the following universal relation
\beq
\lambda n^\star _f {T(R) \over C_2} =1 \, . \label{result}
\eeq
The above relation \eqref{result} is argued to hold for theories based
 on the gauge groups $SU(N)$, $SO(N)$ and $Sp(2N)$ with $N_f$ massless fermions in either the fundamental, adjoint, symmetric or anti-symmetric representation. $\lambda$ is a universal constant (presumably $\lambda \approx 1/4$). $T(R)$ is defined by ${\rm tr} \, T^a T^b = T(R) \delta ^{ab}$. $C_2$ denotes the quadratic Casimir of the adjoint representation $T_{\rm adj.}^a T_{\rm adj.}^a = C_2 1$ ($C_2 = T({\rm adj.})$). $n^\star_f$ denotes the number of Weyl (or Majorana) fermions. Throughout the paper we will use the notation $N_f$ for Dirac fermions (note that $n_f = 2N_f$).

The basic idea behind the present work is to expand the fermionic determinant in powers of $N_f$. The zeroth order correspond to the quenched theory, where the theory is confining. We argue that the series converges as long as $N_f < N^\star _f$ and we use it to derive our result \eqref{result}. We repeat the derivation for various representations and gauge groups and show that the coefficient $\lambda$ is universal, i.e. it does not depend on the representation or the gauge group. In $SU(N)$ theories $\lambda \sim g_{\rm st} N_c$, where $g_{\rm st}$ is the three glueballs coupling of pure Yang-Mills theory (or the string coupling of the string dual).

The paper is organized as follows: in section 2 we introduce our idea and use it for the case of $SU(N_c)$ Yang-Mills theory coupled to $N_f$ massless fundamental flavors. In section 3 we generalize our discussion to other representations and other gauge groups. In section 4 we make a crude estimate of $\lambda$ by using the knowledge of $N_f^\star$ in SQCD. Finally, in section 5 we discuss our results and compare them to other approaches.

\section{The conformal window in $SU(N_c)$ Yang-Mills with $N_f$ fundamental fermions}

Our discussion is based on an earlier paper \cite{Armoni:2008jy}. Consider a calculation of the expectation value of a large circular Wilson loop in multi flavor QCD (by large we mean $\Lambda _{\rm QCD} R \gg 1$). In the path integral formalism it can be written, after integration over the fermions, as follows
\beq 
\langle {\cal W} \rangle _{\rm QCD} = \frac{1}{\cal Z} \int DA_\mu \, {\cal W} \, \exp \left( -S_{\rm YM} \right ) \left ( \det  i \slsh \!D  \right ) ^{N_f} \, . \label{part1}
\eeq   

Let us use the worldline formalism \cite{Strassler:1992zr} in order to express the fermionic determinant in terms of Wilson loops. The fermionic determinant is related to the Wilson loop as follows
\beq
\left ( \det i\slsh \!D  \right )^{N_f}  = \exp N_f \Gamma [A] \, ,
\eeq
where
\bea
\label{wlineint}
 \Gamma [A] &=&
-{1\over 2} \int _0 ^\infty {dT \over T}
\nonumber\\[3mm]
 &\times&
\int {\cal D} x {\cal D}\psi
\, \exp
\left\{ -\int _{\epsilon} ^T d\tau \, \left ( {1\over 2} \dot x ^\mu \dot x ^\mu + {1\over
2} \psi ^\mu \dot \psi ^\mu \right )\right\}
\nonumber \\[3mm]
 &\times &  {\rm Tr }\,
{\cal P}\exp \left\{   i\int _0 ^T d\tau
\,  \left (A_\mu \dot x^\mu -\frac{1}{2} \psi ^\mu F_{\mu \nu}  \psi ^\nu
\right ) \right\}  \, ,
\eea
with $x^\mu (0)=x^\mu (T)$.
Thus $\Gamma [A]$ is a sum over (super)-Wilson loops. The sum is over contours of all sizes and shapes. The sum can be written schematically as $\Gamma [A] = \sum _{\cal C} w$. In this notation the fermionic determinant is 
\beq
\left ( \det \left (   i\slsh \!D \right ) \right ) ^{N_f} = \exp  N_f \sum _ {\cal C} w = \sum _ n \frac{1}{n!} (N_f)^n \left ( \sum _ {\cal C} w \right )^n \, . \label{fdet}
\eeq
Thus the expectation value of the Wilson loop in QCD is 
\beq 
 \langle {\cal W} \rangle _{\rm QCD}  =  \langle {\cal W} \rangle _{\rm YM} + N_f \sum _{\cal C} \langle {\cal W } w \rangle _{\rm YM} ^{\rm conn.} + N_F^2 \sum _{\cal C} \sum _{\cal C} \langle {\cal W } w w \rangle _{\rm YM} ^{\rm conn.} + ...  \label{expansion1} \, .
\eeq
The above expansion is performed in the full non-perturbative Yang-Mills vacuum.
The $1/n!$ in \eqref{fdet} cancels against a combinatorial $n!$.

\begin{figure}[!ht]
\centerline{\includegraphics[width=9cm]{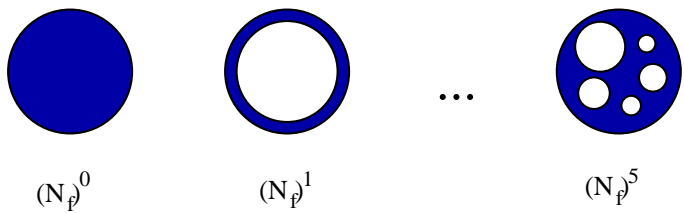}}
\caption{\footnotesize The expectation value of a large circular Wilson loop in QCD. The $O(N_f ^0)$ term gives rise to an area law. Higher order terms in the expansion give rise to a perimeter law, as the fermion loops create holes in the worldsheet.} \label{fig1}
\end{figure}

The first term in the expansion \eqref{expansion1} is given by (see fig. \eqref{fig1})
\beq
 \langle {\cal W} \rangle _{\rm YM} = N_c \exp -\sigma A \,
\eeq
since the YM theory confines. The second term in \eqref{expansion1} is dominated by a Wilson loop $w$ which creates the largest hole in the worldsheet of ${\cal W}$, except a narrow boundary given by a minimal distance $l$ (a UV cut-off), see figure \eqref{fig1}. It can be demonstrated either by the lattice strong coupling expansion or by the AdS/CFT. In particular in the AdS/CFT framework, the two point function $ \langle {\cal W } w \rangle _{\rm YM} $ is given by the Nambu-Goto action
\beq
 \langle {\cal W } w \rangle _{\rm YM} = C \exp - I_{\rm N.G.} \,.
\eeq
The Wilson loop $w$ that minimizes the Nambu-Goto action, is depicted in figure \eqref{fig1}, see ref.\cite{Armoni:2008jy} for a detailed discussion. The result is a perimeter law
\beq
 \langle {\cal W } w \rangle _{\rm YM} = C_1 N_f \exp -\mu P \,.
\eeq

Higher order terms in the expansion will also create holes and result in a perimeter law. They are proportional to $(N_f)^n$ (where $n$ is the order of the expansion). To be more precise we argue that the $n^{\rm th}$ term in the expansion is given by
\beq
 \langle {\cal W } w ... w \rangle _{\rm YM} = C_n N_c \left ({N_f \over N_c} \right ) ^n  \exp -\mu P \, , 
\eeq
namely that the expansion is in powers of ${N_f \over N_c}$. In order to obtain an intuition about the above expansion let us assume that in passing from order $n$ to an order $n+1$ we need to add a {\it small} Wilson loop, which is given by 
\beq
w \sim g_{\rm YM}^2 {\rm tr} \, F^2 a^2 \sim  {\lambda \over N_c} ( {\rm tr}\, F^2 a^2)\, \label{pert} ,
\eeq
where $\lambda$ is the 't Hooft coupling and $a$ is the area of the loop. The assumption that we need to add a small Wilson loop is not valid. However we will soon argue that it captures the right dependence on $N_f$ and $N_c$. 

\begin{figure}[!ht]
\centerline{\includegraphics[width=9cm]{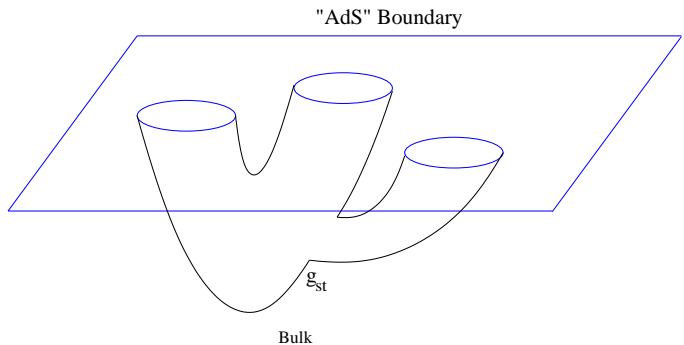}}
\caption{\footnotesize An example of a calculation of Wilson loops correlation function: a calculation of $\langle www \rangle _{\rm conn.}$. The bulk is a typical closed string theory diagram with one string interaction.} \label{three}
\end{figure}

In the AdS/CFT framework adding a Wilson loop to a diagram which consists of $n$ Wilson loops will result in a factor of $g_{\rm st} = \lambda / N_c$, since we need to connect the $(n+1)^{\rm th}$ Wilson loop worldsheet to the bulk worldsheet. An example is given in figure \eqref{three}\footnote{Figure \eqref{three} suggests the existence of bulk non-planar $g_{\rm st}^2$ corrections. These corrections will renormalize the QCD string tension.}. We thus conclude that the $n^{\rm th}$ term in the expansion is given by
 \beq
   N_c \left (\lambda {N_f \over N_c} \right ) ^n  \exp -\mu P \, . 
\eeq
Thus $\lambda$ is a constant proportional to the 't Hooft coupling of the strongly coupled pure Yang-Mills theory. 

The expansion in terms of Wilson loops (or in powers of $N_f$) makes sense provided that it has a finite, but non-zero, radius of convergence. Although it is not necessary, we assume that the series is geometric. Our {\it assumption} that it forms a geometrical series namely that $C_n = \lambda ^n$ is supported by the string expansion and also by Veneziano's topological expansion \cite{Veneziano:1976wm}. The expansion will break down at a critical $N_f^\star$ when
\beq
\lambda {N^\star_f \over N_c } =1 \,. \label{breaks}
\eeq 
The expansion of QCD observable around the confining Yang-Mills vacuum makes sense as long as the theory is in the confining phase. In particular, the zeroth order term $\exp -\sigma A$, corresponds to a mass gap in the gluonic sector of the theory. Such a term is not acceptable inside the conformal window and hence we argue that eq.\eqref{breaks} determines the lower boundary of the conformal window. 

\section{Other representations and gauge groups}

The generalization to other representations is straightforward. The perturbative expansion of a small Wilson loop \eqref{pert} suggests that when the fundamental representation is replaced by a different representation $R$, eq.\eqref{breaks} is replaced by
 \beq
\lambda_R n^\star_f {T(R) \over N_c } =1 \,. \label{breaks2} 
\eeq
a priori , the coefficient $\lambda _R$ depends on the representation of the matter field. 

\begin{figure}[!ht]
\centerline{\includegraphics[width=2cm]{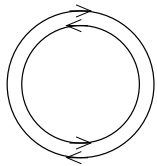}}
\caption{\footnotesize A Wilson loop in the adjoint representation.} \label{adjoint}
\end{figure}

Let us start with the adjoint representation. Adjoint fields will not screen the fundamental loop and therefore will not create a hole in the worldsheet. Instead each term in the expansion will contribute $\exp -\sigma A$, since worldsheets with a higher area will be exponentially suppressed. In the 't Hooft double line notation, the adjoint loop contains two lines, see figure \eqref{adjoint}. When we connect an adjoint Wilson loop to a fundamental loop, we actually connect only one of the lines\footnote{Non-planar graphs where both lines of the adjoint loop are connected to other Wilson loops are exponentially suppressed with respect to graphs where only one line is connected to other Wilson loops.}, while the second line provides an extra factor of $N_c$. In particular, when the fermions are in the adjoint representation, $\langle {\cal W}w \rangle \sim N_c n_f$. Higher order corrections are of order ${\cal O}((g_{\rm st} N_c n_f)^n)$. Thus 
\beq
\lambda_{\rm adj.} n^\star _f =1 \, .
\eeq

\begin{figure}[!ht]
\centerline{\includegraphics[width=6cm]{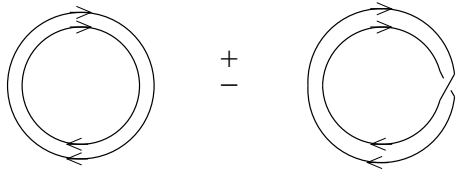}}
\caption{\footnotesize A Wilson loop in the two-index symmetric or antisymmetric representation.} \label{symm}
\end{figure}

The generalization to matter in the two-index symmetric (or two-index antisymmetric) is as follows: by using the decomposition in terms of fundamental Wilson loops ${\rm tr}\, u_{\rm symm.} ={1\over 2} \left ( {\rm tr}\, u_f {\rm tr}\, u _f + {\rm tr}\, u_f^2 \right )$ (or  ${\rm tr}\, u_{\rm Asymm.} ={1\over 2} \left ({\rm tr}\, u_f {\rm tr}\, u _f - {\rm tr}\, u_f^2 \right )$) we learn that we need to sum two contributions: the first is similar to the adjoint (however, in the 't Hooft notation the two lines admit the same orientation) and the second contains one line that winds twice, see figure \eqref{symm}. The Nambu-Goto action for both contributions is identical, but one is ${\cal O}(N_c)$ while the other is ${\cal O}(1)$, see ref.\cite{Armoni:2006ri}. The sum of the two contributions is ${1\over 2}(N_c+2)$ (or ${1\over 2}(N_c-2)$). This assertion is consistent with the fact that for $SU(2)$ the antisymmetric representation is identical to the singlet. Since the singlet decouples from the Yang-Mills theory, a factor $N_c-2$ is anticipated for the antisymmetric Wilson loop. We thus obtain,
\beq
\lambda_{\rm symm/Asymm} N^\star _f {N_c \pm 2 \over N_c} =1 \, .
\eeq

Large $N_c$ equivalence \cite{equivalence} between a theory with $N_f$ adjoint Majorana fermions and $N_f$ symmetric (or antisymmetric) Dirac fermions yields
$\lambda _{\rm adj.} = \lambda _{\rm symm/Asymm}$. Moreover, for $SU(3)$ a Dirac fermion in the fundamental representation is equivalent to a Dirac fermion in the antisymmetric representation. Hence $ \lambda _{\rm symm/Asymm} = \lambda _f$. These observations lead to the relation
 \beq
\lambda n^\star_f {T(R) \over N_c } =1 \, , \label{breaks3} 
\eeq
for $SU(N_c)$ theories with matter in either the fundamental/adjoint/symmetric or antisymmetric representation.

The generalization to $SO(N_c)$ or $Sp(2N_c)$ gauge group is achieved by noticing that the string coupling is related to $N_c$ by 
\beq
g_{\rm st} = {\lambda _{\rm SO/Sp} \over C_2 } \, ,
\eeq
where $C_2$ is the quadratic Casimir of the adjoint representation of the $SO$/$Sp$ group, see ref.\cite{Sinha:2000ap} for a recent discussion. Therefore, when passing from the $n^{\rm th}$ order to the next order, we ``pay'' by a factor of $n_f {\lambda _{\rm SO/Sp} \over C_2 }$.  

Note also that planar equivalence between $SU(N_c)$ and $SO/Sp$ theories leads to $\lambda _{\rm SU} = \lambda _{\rm SO/Sp}$. We therefore propose \eqref{result} as a universal relation, with the same universal constant $\lambda$ for $SU$, $SO$ and $Sp$ theories.

We end this section by listing the values of $T(R)$ for the cases under consideration, see table (1) below.
  
\begin{table}[!ht]
\begin{center}
\begin{tabular}{|c|c|c|}
\hline
                Group & Rep. & $T(R)$ \\
\hline
 $SU(N_c)$ & $\fund $ & $\smallfrac{1}{2}$ \\
\hline
 $SU(N_c)$ & ${\rm adj.} $ & $N_c$ \\
\hline
 $SU(N_c )$ & $ \symm $ & $\smallfrac{1}{2}(N_c+2) $ \\
\hline
 $SU(N_c )$ & $ \asymm $ & $\smallfrac{1}{2}(N_c-2) $ \\
\hline
\hline
 $SO(N_c)$ & $\fund $ & $ 1 $ \\
\hline
 $SO(N_c)$ & ${\rm adj.} $ & $N_c-2 $ \\
\hline
 $SO(N_c )$ & $ \symm $ & $ N_c+2 $ \\
\hline
\hline
 $Sp(2N_c)$ & $\fund $ & $ \smallfrac{1}{2}$ \\
\hline
 $Sp(2N_c)$ & ${\rm adj.} $ & $N_c+1$ \\
\hline
 $Sp(2N_c )$ & $ \asymm $ & $ N_c-1 $ \\
\hline

\end{tabular}
\label{TC}
\caption{$T(R)$ for various representations of $SU(N)$, $SO(N)$ and $Sp(2N)$.}
\end{center}
\end{table}

\section{An estimate of $\lambda$ from SQCD}

In SQCD $N_f^\star = {3 \over 2}N_c$. Let us use it to estimate the value of $N_f^\star$ in non-supersymmetric QCD.

In order to generalize our discussion to the SQCD case, we need to incorporate scalars. The worldline formalism for scalars is, in fact, simpler, since scalars carry no spin. In this case there are no worldline fermions and the Wilson loop is purely bosonic \cite{Strassler:1992zr}
\beq
\label{wlineint2}
 \Gamma [A] =
 \int _0 ^\infty {dT \over T}
\int {\cal D} x \, \exp -\int _{\epsilon} ^T d\tau \, \left ( {1\over 2} \dot x ^\mu \dot x ^\mu \right )
 \times   {\rm Tr }\,
{\cal P}\exp  i\int _0 ^T d\tau
\,  \left (A_\mu \dot x^\mu  \right )  \, .
\eeq

In our discussion we consider the coupling of a large Wilson loop ${\cal W}$ to Wilson loops which are generated by the scalar or the fermionic determinant. Our consideration is semiclassical, namely we assume a saddle-point configuration which dominates the path integral. Let us argue that we can neglect the fermion spin and that in the semiclassical approximation we can put scalars and fermions on equal footing. The reason is that the classical action for the worldline fermions is quadratic (for simplicity we write down the action when the fermions are coupled to an Abelian gauge field)
\beq
S_f = \int d \tau \, \left ( {1\over 2} \psi ^\mu \dot \psi ^\mu -\frac{1}{2} \psi ^\mu F_{\mu \nu}  \psi ^\nu  \right ) \, ,
\eeq
and therefore $\psi ^\mu =0$ is a solution of the worldline-fermions equation of motion $\delta S / \delta \psi =0$. Thus the saddle point solution for the correlation function of ${\cal W}$ with either a super-Wilson loop or a bosonic Wilson loop is identical. For this reason, we assume that we can neglect the spin of the fermion and we estimate the contribution of a scalar to be identical to the contribution of a fermion. 

SQCD consists of $N_f$ fundamental fermions, $N_f$ fundamental scalars and one adjoint fermion. Under the estimate that scalars and fermions contribute in equal weight we obtain
\beq
{\lambda \over N_c} \sum _R  n _f(R) T(R)  = \lambda \left (2{N^\star _f \over N_c} +1 \right) \approx 1 \, . \label{sqcd}
\eeq
Substituting $N^\star _f / N_c =3/2$ in the above equation, we obtain
\beq
\lambda \approx {1\over 4} \, , \label{estimate}
\eeq
hence in QCD (or QCD-like theories) we estimate the lower boundary of the
 conformal window by
\beq
 n^\star _f {T({\rm matter}) \over C_2 } \approx 4 \, . \label{estimate2}
\eeq
We would like to stress that the above estimate is not fully justified: the SQCD Lagrangian consists of a Yukawa interaction term between the squark the quark and the gluino. In our discussion we have ignored it --- without a proper justification. It would be nice if there exists a limit in which the Yukawa term could be ignored and our estimate is justified.

\section{Discussion}

In this paper we argued that the conformal window in gauge theories respects a universal relation \eqref{result}. The relation is general and it applies to various gauge groups and various representations. It contains a universal unknown constant $\lambda $ which is essentially the ``string coupling'' (times $N_c$ in $SU(N_c)$ theories), namely the strength of the three glubeballs interaction. The constant $\lambda$ can be evaluated in a lattice simulation of pure Yang-Mills theory by considering a three point function of Wilson loops.

The relation \eqref{result} holds in supersymmetric theories, since the NSVZ beta function admits a non-trivial zero when $n_f T(R)/ C_2(G) = 3/2$ \cite{NSVZ} (here $n_f$ counts the number of chiral superfields). Interestingly, an NSVZ-inspired beta function was proposed in \cite{Ryttov:2007cx,Sannino:2009aw}. An outcome of this proposal is our result \eqref{result} with a universal constant $\lambda = 4/11$.

The relation \eqref{result}, for $SU(N_c)$ theories, was also obtained in \cite{Poppitz:2009uq}, namely the dependence of the representations is proportional to $T(R)/N_c$, with $\lambda =1/4$, except for the fundamental representation, where $\lambda =2/5$. 

Our estimate for QCD with $N_f$ fundamental Dirac fermions is $N_f^\star \approx 4N_c$. It is in agreement with \cite{Appelquist:1996dq,Miransky:1996pd,Appelquist:1999hr}, but it differs from other approaches. We wish to stress that all the estimates of $N_f^\star$, {\it including ours}, for non-supersymmetric theories on $R^4$ rely on assumptions that may, or may not, be valid.

Lattice simulations of QCD with matter in various representations are carried out currently \cite{lattice}. After the dust settles, those simulations will confirm or rule out the relation \eqref{result}. 

\vskip 1cm

{\it Acknowledgements.} I wish to thank P. Kumar, T. Hollowood, C. Nunez, A. Patella, M. Piai, M. Shifman and M. Unsal for fruitful discussions
and comments. I am supported by the STFC advanced fellowship award.

%%%%%%%%%%%%%%%%%%%%%%%%%%%%%%%%%%%%%%%%%%%%%%%%%%%%%%%%%%%%%%%%%%%5

\end{document}